\documentclass[twocolumn,showpacs,preprintnumbers,amsmath,amssymb,prapplied,superscriptaddress]{revtex4-1}
\usepackage{graphicx}
\usepackage{dcolumn}
\usepackage{bm}
\usepackage{color}
\usepackage{braket}
\begin{document}

\preprint{APS/123-QED}

\title{Scalability of all-optical neural networks based on spatial light modulators}

\author{Ying Zuo} \thanks{These authors contributed equally to this work.}
\affiliation{Department of Physics, The Hong Kong University of Science and Technology, Clear Water Bay, Kowloon, Hong Kong, China}

\author{Zhao Yujun} \thanks{These authors contributed equally to this work.}
\affiliation{Department of Physics, The Hong Kong University of Science and Technology, Clear Water Bay, Kowloon, Hong Kong, China}

\author{You-Chiuan Chen}
\affiliation{Department of Physics, The Hong Kong University of Science and Technology, Clear Water Bay, Kowloon, Hong Kong, China}

\author{Shengwang Du}\email{dusw@utdallas.edu}
\affiliation{Department of Physics, The University of Texas at Dallas, Richardson, Texas 75080, USA}

\author{Junwei Liu} \email{liuj@ust.hk}
\affiliation{Department of Physics, The Hong Kong University of Science and Technology, Clear Water Bay, Kowloon, Hong Kong, China}

\date{\today}

\begin{abstract}
Optical implementation of artificial neural networks has been attracting great attention due to its potential in parallel computation at speed of light. Although all-optical deep neural networks (AODNNs) with a few neurons have been experimentally demonstrated with acceptable errors recently, the feasibility of large scale AODNNs remains unknown because error might accumulate inevitably with increasing number of neurons and connections. Here, we demonstrate a scalable AODNN with programmable linear operations and tunable nonlinear activation functions. We verify its scalability by measuring and analyzing errors propagating from a single neuron to the entire network. The feasibility of AODNNs is further confirmed by recognizing handwritten digits and fashions respectively. 
\end{abstract}

\maketitle

\section{Introduction}

In the last decades, artificial neural networks have grown into one of the most disruptive technologies \cite{jordan2015machine,butler2018machine,alpaydin2020introduction} and found success in both practical applications, such as computer vision, speech recognition and natural language processing\cite{ciregan2012multi,graves2013speech,goldberg2017neural}, and fundamental studies such as particle physics, condensed matter physics and material science \cite{lAIMaterialDiscovery,carrasquilla2017machine,carleo2017solving,liu2017,huang2017accelerated,liu2017femion,radovic2018machine, shen2018self,torlai2018neural,carleo2019machine,palmieri2020experimental}. The power of an artificial neural network comes from its large amount of artificial neurons and their intensive interconnections, which require large memory size, computational complexity and energy consumption \cite{Merolla668}. Due to the light wave nature, all-optical deep neural networks (AODNNs) hold a great promise in high-speed parallel computation with low energy consumption\cite{jutamulia1996overview,abu1987optical,larger2012photonic,duport2012all,cheng2019optical,jiao2019optical,sui2020review}. AODNNs have not been implemented experimentally until recently with the breakthroughs in realizing optical nonlinear activation functions using phase-change materials \cite{feldmann2019all} and electromagnetically induced transparency (EIT) atomic medium \cite{zuo2019all}, which explicitly remove all the principle restrictions of constructing a general-purpose AODNN. However, any of these demonstrated AODNNs has no more than 22 neurons and the scalability remains a question.

Different from a universal digital electronic implementation of artificial neural networks, an optical neural network is usually designed and built for a specific given task \cite{woods2012photonic, ShenNP2017, lin2018all}, and one has to rebuild a completely new optical neural network even for a slightly different task which shares similar structure. In addition, the error from the imperfection of optical signals and components may accumulate exponentially with the number of neurons to completely ruin the performance in a large optical neural network even though in general large-size neural networks have large tolerance for random local error from an individual neuron \cite{sung2015resiliency}.

In this work, we demonstrate a scalable AODNN based on free-space Fourier optics and EIT nonlinear activation functions. The programmable linear transformations in our AODNNs are realized using spatial light modulators\cite{lu1990theory} (SLMs) and optical lenses. We experimentally show that local random errors can be reduced by increasing the area of SLM ($S$) per neuron. Moreover the total error in the linear transformation only accumulates with the deviation proportional to square root of the number of neurons ($M$). In addition, the nonlinear activation functions realized with EIT in cold atoms are spatially separated, thus their errors are independent and accumulates similarly to the linear transformation (with the deviation proportional to square root of their number).
Therefore this AODNN can be scaled up to large size since effect of this kind of slowly accumulated errors can be easily compensated by increasing the number of hidden neurons\cite{sung2015resiliency}.
We confirm the feasibility and programmability by experimentally constructing a fully-connected AODNN with 174 optical neurons, and apply it to recognize handwritten digits and fashions, with classification rates of 81.8\% and 71.3\%, respectively. Furthermore, we clearly show its scalability using accurate numerical simulations, where the classification rates indeed increase with the number of optical neurons even with large random error.
\begin{figure*}
\includegraphics[width=\linewidth]{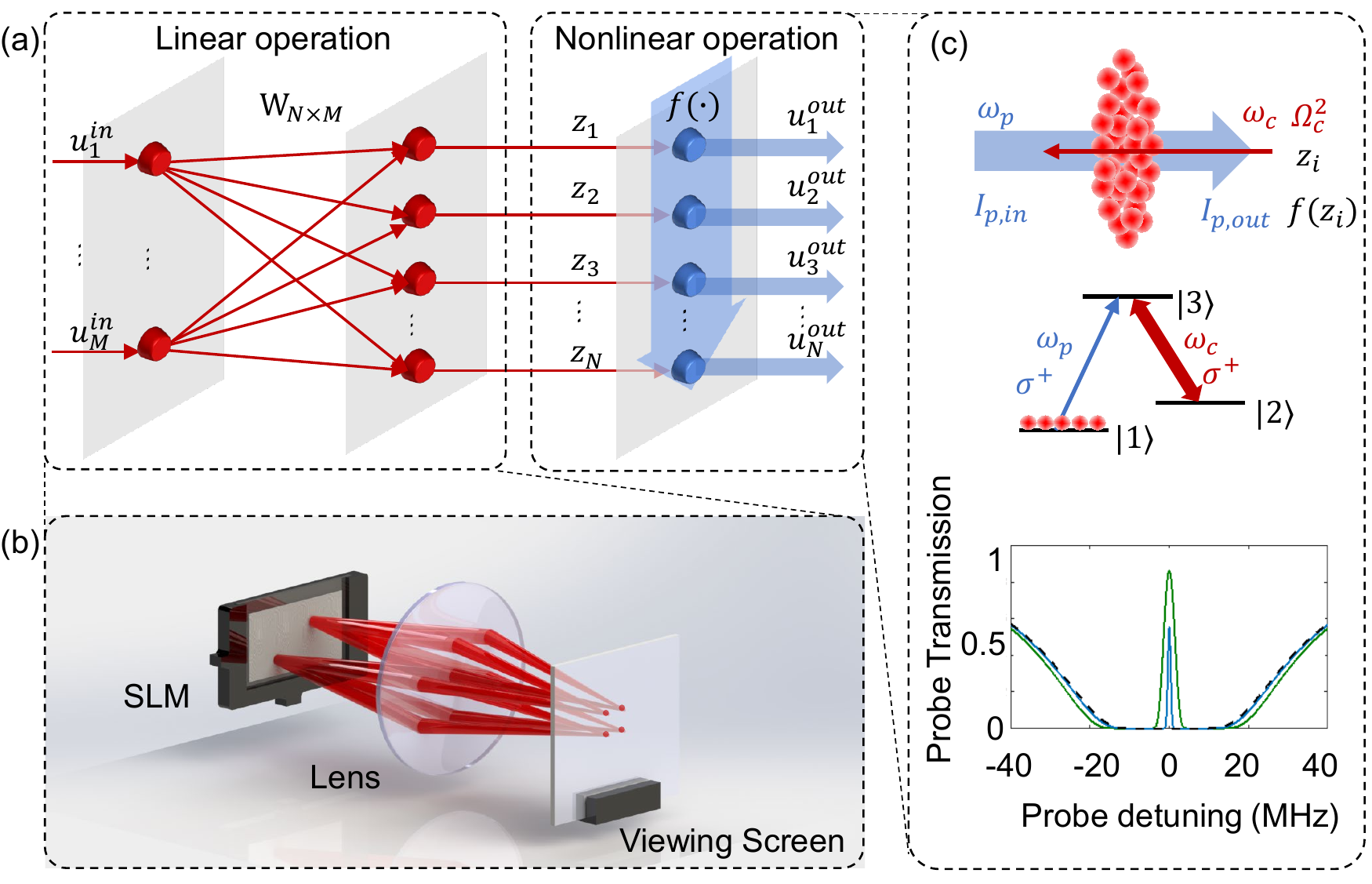}
\centering
\caption{\label{fig:figure1}Optical neurons and connections. (a) Schematics of optical neurons and connections between two adjacent layers in an AODNN. The N nodes $u^{in}_j$ form the input layer. $\mathbf{W_{N\times M}}$ is the $N\times M$ linear linear matrix connection. $f_i(z_i)$ are M outputs after the nonlinear activation functions $f_i(\bullet)$. (b) Optical implementation of linear matrix operation $\mathbf{W_{N\times M}}$. A SLM, placed on the back focal plane of an optical lens, splits each input optical nodes into multiple directions. The lens performs Fourier transform and sum the light beams along the same directions onto a spot on its front focal plane. (c) EIT nonlinear optical activation function with a three-level atomic medium (top panel) and EIT transmission spectrum of the probe beam (bottom panel). The transmission of the probe beam power $I_{p,in}$ is nonlinearly controlled by the power $I_{c}$ of the control light. The relevant $^{85}$Rb atomic energy level are $\ket{1}=\ket{5^2 S_{1/2},F=2}$, $\ket{2}=\ket{5^2 S_{1/2}, F=3}$ and $\ket{3}=\ket{5^2 P_{1/2}, F=3}$. The black dashed line in the bottom panel is obtained without the control beam and shows maximum absorption on resonance at zero probe detuning. The solid lines (blue and green) are two typical EIT spectra with different control light power.}
\end{figure*}
\section{Optical neurons and interlayer connections.}

Fig.\ref{fig:figure1} shows the schematics of optical neurons and their interconnections between two adjacent layers. As the building block of an artificial neural network, neurons receive input signals $u_j^{in}$, conduct a linear transformation $z_i=\sum_j W_{ij} u_j^{in}$, pass the results to the nonlinear activation function $f_i(\bullet)$ and produce the output signals $u_i^{out}=f_i(z_i)$ as shown in Fig.\ref{fig:figure1}(a).
In our AODNN, signals are encoded in the power of light, and computations are conducted through light propagation, diffraction, and interference. In detail, we perform a linear transformation using a SLM and a Fourier lens, as shown in Fig.~\ref{fig:figure1}(b). The input light beams ($u_1^{in}\sim u_n^{in}$) are incident on different parts of the SLM, which is placed at the back focal plane of the lens. Through controlling the multiple phase gradings in a certain area of the SLM, the input beam array $u^{in}_j$ are diffracted to different directions with powers $W_{ij} u^{in}_j$. Then, these weighted beams from different neurons are focus onto the same spot on the front focal plane of the lens and complete the linear summation $z_i = \sum_j W_{ij} u^{in}_j$. 
It is worth noting that the computation time here is independent of the number of neurons and interlayer connections but determined by the propagation time of light from the SLM to the front focal plane of the lens. The wave nature of light makes it possible to perform linear transformation with massive parallelism at the speed of light. Moreover, different from many other systems, the layer structure, including the number of neurons and the weight matrix, in our system are fully programmable.

The optical nonlinear activation functions are realized using EIT ~\cite{EITHarris,fleischhauer2005electromagnetically} effect in cold $^{85}$Rb atoms prepared in a two-dimensional (2D) magneto-optical trap (MOT) \cite{metcalf2003laser, 2DMOTRSI2012}. As shown in top panel of Fig.~\ref{fig:figure1}(c), a pair of counter-propagating probe ($\omega_p$) and control ($\omega_c$) beams are shined on the atomic medium. The relevant $^{85}$Rb atomic energy levels are labelled as $\ket{1}$, $\ket{2}$ and $\ket{3}$, and the atoms are in the state $\ket{1}$. The circularly ($\sigma^+$) polarized probe laser is on resonance to the transition $\ket{1}\to \ket{3}$, and the circularly ($\sigma^+$) polarized control laser is on resonance to the transition $\ket{2}\to \ket{3}$. Without the control light, the atomic medium is opaque to the probe beam due to its on-resonance absorption. In presence of the control beam, the atomic medium becomes transparent and the probe transmission depends on the power of the control light. The bottom panel of Fig.~\ref{fig:figure1}(c) shows typical probe transmission spectrums with different control light powers. The nonlinear dependency of the output probe power ($I_p^{out}$) on the control laser power ($I_c$) is described as $I_{p}^{out}=f(I_c)=I_{p}^{in}e^{-OD \frac{4\gamma_{12}\gamma_{13}}{\Omega_c^2+4\gamma_{12}\gamma_{13}} }$, where $I_{p}^{in}$ is the input probe beam power. $\Omega_c$ is the Rabi frequency of the control beam and $\Omega_c^2$ is proportional to control beam intensity $I_c$. Here, $\gamma_{13}=2\pi\times3$ MHz is fixed and determined by the spontaneous emission of the excited state $\ket{3}$. The ground-state dephasing rate $\gamma_{12}$ can be engineered by applying external magnetic field. $OD$ is the atomic optical depth on the probe transition and can be varied by changing the atomic density. In our optical neuron, the control light works as the input signal to its nonlinear activation function and the probe light output is the output signal of the activation function. The EIT optical nonlinear activation function can be tuned by manipulating the cold atom parameters $OD$ and $\gamma_{12}$, and they are also independent for different neurons. 
Moreover, the energy cost here is very small. Under the experimental condition that $OD\approx 2$ and $\gamma_12 \approx 5$, the activation function shows strong nonlinearity with coupling power to be less than $1mW$ with the diameter of beam to be $100\mu m$.
\begin{figure*}
\includegraphics[width=\textwidth]{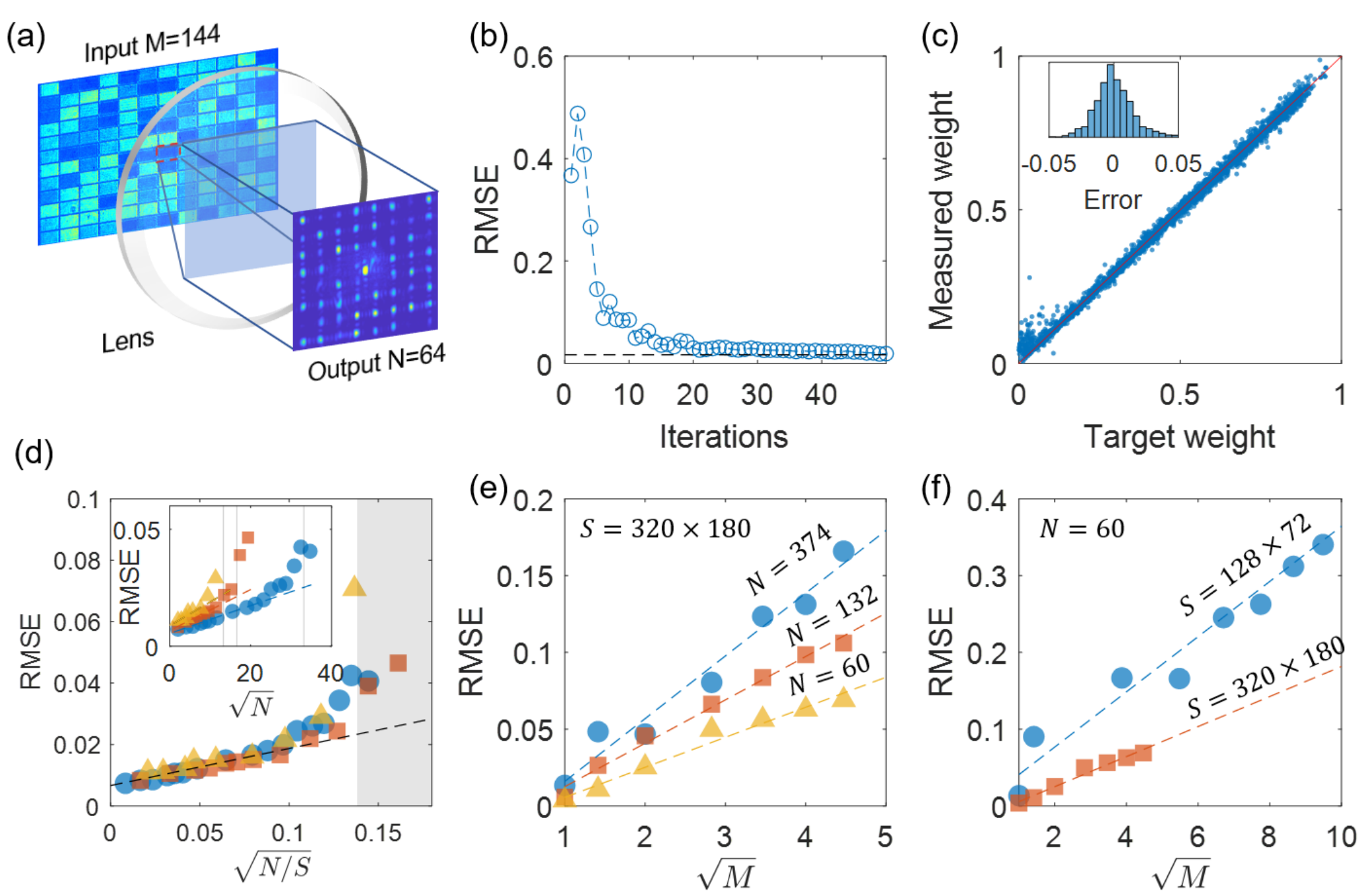}
\centering
\caption{\label{fig:figure2} \textbf{Scalability of linear operation $\mathbf{W}_{N\times M}$.} \textbf{a}, Typical images of input nodes vector $\mathbf{u_M}^{in}$ and the weighted output $W_{ij}u^{in}_j$ from an individual input node $u^{in}_j$. For the purpose of illustration, we take here $N=64$ and $M=144$. \textbf{b}, Error during the GSW feedback iteration process for configuring the weights $W_{ij}$. The data are taken with $S=320 \times 180$ and $N=400$. \textbf{c}, The measured weights v.s. the target weights. The inset histogram shows the error distribution. \textbf{d} is the dependency of error on $\sqrt{N/S}$. The yellow triangles, red squares and blue circles are the experimental RMSE data of $S=128\times72$, $S=160\times90$ and $S=320\times180$ accordingly. The dashed lines are the fitting curves of each case. \textbf{e} and \textbf{f} are the dependency of error on $\sqrt{M}$ with fixed $S$ and $N$, respectively.}
\end{figure*}
\section{Error and scalability}

Clearly, the scalability of our AODNNs is determined by accuracy of the linear interlayer connections $\mathbf{z}_N =\mathbf{W}_{N\times M} \mathbf{u}^{in}_M$, where $\mathbf{z}_N $ ($\mathbf{u}^{in}_M$) is a $N$ ($M$) dimensional vector, and $\mathbf{W}_{N\times M} $ is an $N\times M$ matrix. In our optical implementation, such a linear transformation is divided into two steps: (1) Weighted multiplication by the SLM: for any given $j\in [1, M]$ and $i\in [1, N]$, $y_i^j=W_{ij} u^{in}_j$; and (2) summation by the lens: $z_i=\sum_{j=1}^M y_i^j$. $M$ and $N$ are the number of input and output nodes. As illustrated in Fig.~\ref{fig:figure1}(b), the SLM is divided into $M$ sections and the beam on each section is split into $N$ directions. Fig.~\ref{fig:figure2}(a) displays typical optical intensity distribution patterns of $M=144$ input nodes on the SLM and $N=64$ output nodes split from one of the input nodes. Obviously, the $M$ input sections ($\mathbf{u}^{in}$) on the SLM are independent. For a given input node $u^{in}_j$, its splitting into $N$ weighted outputs, i.e. $y_i^j=W_{ij} u^{in}_j$ are handled by the same section area $S$ of SLM and inevitably entangled with each other.

We use the weighted Gerchberg-Saxton (GSW) algorithm\cite{matsumoto2012high,kim2019large,nogrette2014single} to obtain an appropriate phase pattern applied on SLM for achieving target weights $W_{i j}$. At each iteration of GSW algorithm, we measure powers of output spots and compute the root-mean-square-error (RMSE) between measured weights and target weights, which is taken as feedback for the next iteration (more details are shown in Appendix. A). A typical example of iteration process is shown in Fig.~\ref{fig:figure2}(b), where the RMSE rapidly decreases within 20 iteration steps and usually converges in less than 50 steps. To make our test experiment fair, we randomly choose the value of weights between 0 and 1. The test result for $N=400$ and $S=320\times 180$ is shown in Fig.~\ref{fig:figure2}(c). Overall, the accuracy is very high with a small error, and the error is slightly larger for target weights around zero because it is difficult to reduce the light power to zero due to the diffraction from the adjacent beams. As shown in the insert of Fig.~\ref{fig:figure2}(c), the absolute error follows a normal distribution centered at $0$ with a standard deviation of $0.019$, which indicates the random nature of the error and quantitatively show the high accuracy of the linear transformation. We find that the RMSE between measured weights and target weights depends only on $\sqrt{N/S}$, as shown in Fig.~\ref{fig:figure2}(d). For a small $N$, the RMSE is linearly proportional to $\sqrt{N}$ for all different area size $S$ of the input node and the slope of fitted line is proportional to $1/\sqrt{S}$ (more details in \textbf{Supplementary Information S1}). Such a dependency of error on $N$ and $S$ is governed by the nature of Fourier transformation. More details of the theoretically analysis can be found in in \textbf{Supplementary Information S2}. As $N$ becomes very large, the nearest diffracted spots cannot be well separated, which increases the error significantly.  The threshold is indicted by the grey lines in the figure and it clearly increases with $S$ because the area of each diffracted spot decrease with $S$.

We then analyze the error accumulated in step (2) of light power summation by comparing the measured output $\widetilde{z}_{i}$ to the target values $z_{i}=\sum_{j} w_{i j} u_{j}$.
As shown in Fig.~\ref{fig:figure2}(e), for a given $S$, the RMSE$(M,N,S)$ for different $N$ is linearly proportional to $\sqrt{M}$. We also find that for a given $N$, the RMSE$(M,N,S)$ for different $S$ is linearly proportional to $\sqrt{M}$ as shown in Fig.~\ref{fig:figure2}(f) (more details in \textbf{Supplementary Information S1}). Such a $\sqrt{M}$ dependency of RMSE$(M,N,S)$ confirms that the errors from different input nodes are random, independent and uncorrelated, and the total error only accumulates in the stochastic way. Combining with the previous analysis of error dependency on $N$ and $S$, we conclude that the error of optical linear summation increases linearly with $\sqrt{MN/S}$. 
It is worth noting that the linear combination error mentioned is absolute error, which means the relative error decreases with $\sqrt{N/SM}.$

With all errors confirmed to remain local and random, the scalability of our AODNNs is mainly limited by the capacity of linear matrix operation: number of optical neurons and their interconnections between two adjacent layers. 
For a given pixel size ($d\times d$) and number of pixels ($K$) of the SLM, the numerical aperture ($NA$) of the Fourier lens, as well as the light wavelength $\lambda$, our analysis shows the maximum connections, or the linear capacity of a single SLM-lens layer, is determined by $C_L=(M\times N)_{max}=\frac{NA^2 \pi K d^2 }{4\lambda^2}$ (more details in the Appendix B) . For the SLM (HOLOEYE, PLUTO-2) used in this work, we have $d=8\mu$m, $K=1920\times1080$. Together with $NA=0.05$ of the lens and the light wavelength $\lambda=795$ nm, our single SLM-lens layer can accommodate up to $C_L=412,287$ connections. 
In experiment, due to the restricted size of the camera (Hamamatsu C11440-22CU), the maximum connection number we test in Fig.~\ref{fig:figure2}(d) is 46,656. Our system still have potential to be scaled up.

\begin{figure*}
\includegraphics[width=\linewidth]{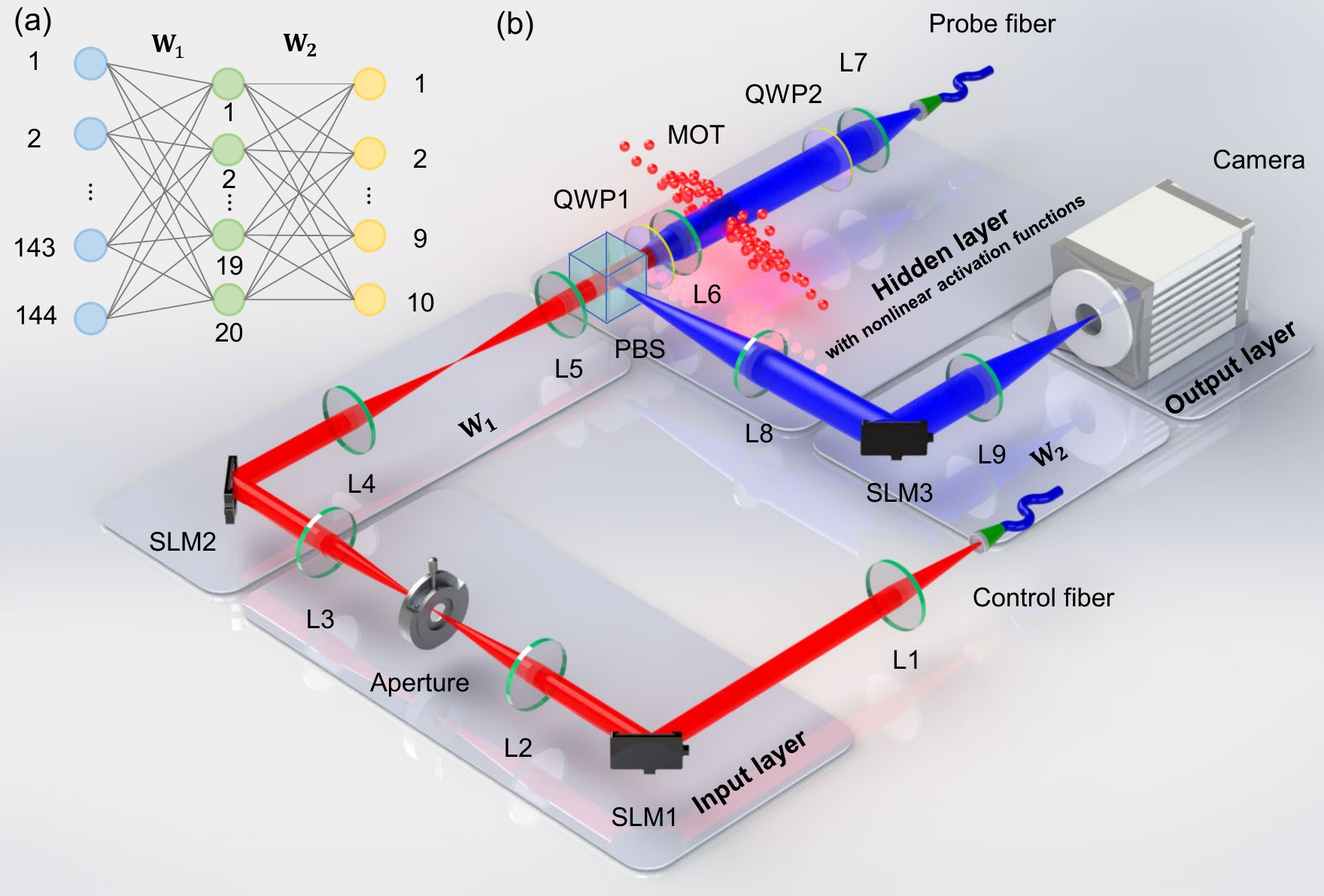}
\centering
\caption{\label{fig:figure3} \textbf{A two-layer AODNN with 144 input neurons, 20 hidden neurons, and 10 output neurons.} \textbf{a}, The corresponding artificial neural network architecture. \textbf{b}, The optical layout of the AODNN. Spatial light modulators: SLM1 (HOLOEY LETO), SLM2(HOLOEYE PLUTO-2), and SLM3(HOLOEY GEAE-2). Camera: Hamamatsu C11440-22CU. PBS: polarization beam splitter. Lenses: L1 ($f=100$ mm), L2 ($f=200$ mm), L3 ($f=250$ mm), L4 ($f=350$ mm), L5 ($f=350$ mm), L6 ($f=50$ mm), L7 ($f=50$ mm), L8 ($f=450$ mm) and L9 ($f=450$ mm). }
\end{figure*}

\begin{figure*}
\includegraphics[width=\linewidth]{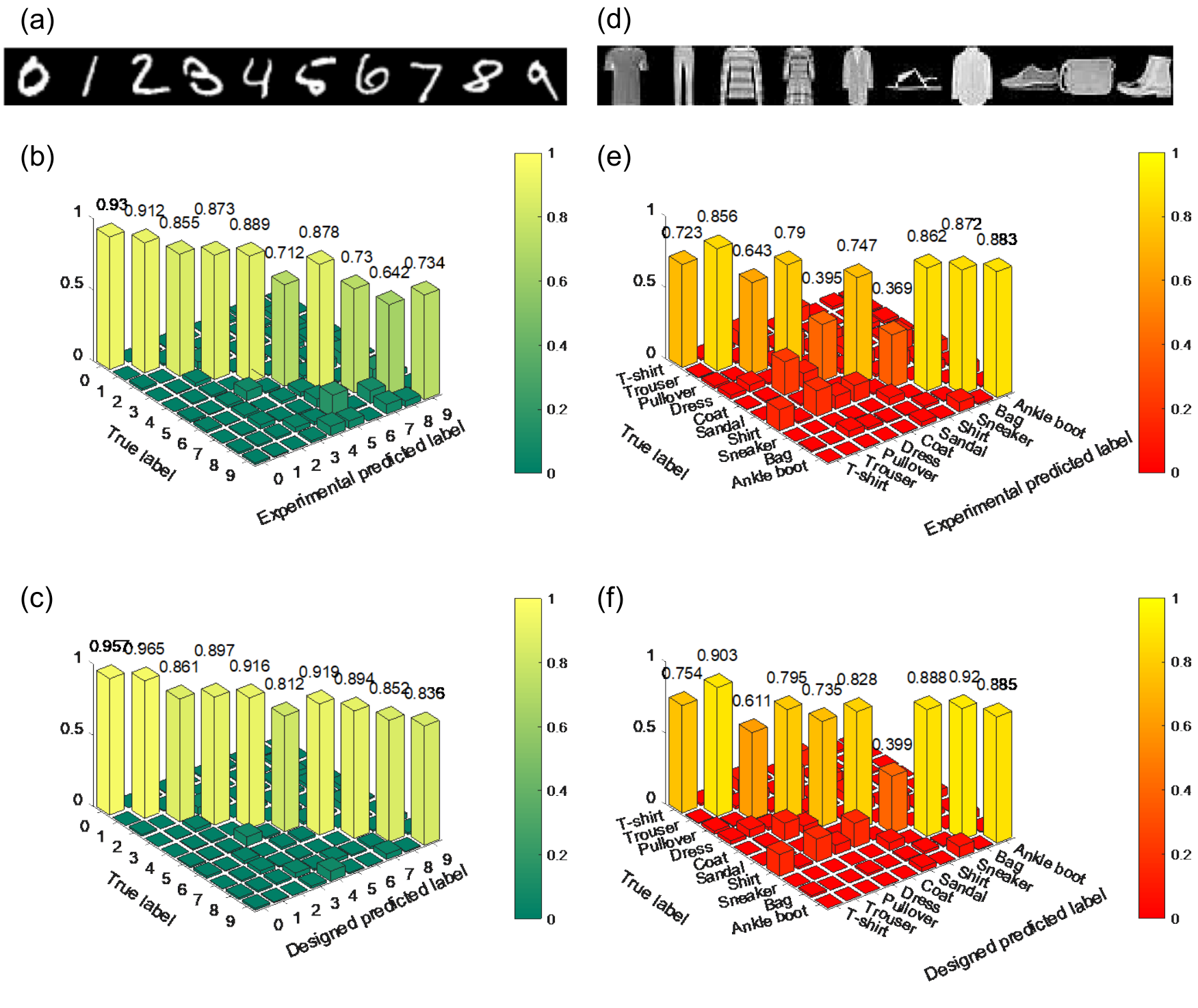}
\centering
\caption{\label{fig:figure4} \textbf{Results of handwritten digit and fashion classifications.} \textbf{a} and \textbf{d} are examples of the test set of 10 handwritten digits and fashion patterns. \textbf{b} and \textbf{e} are the measured normalized confusion matrixes of the AODNN. \textbf{c} and \textbf{f} are the normalized confusion matrixes of computer based neural network with the same structure as the AODNN. Both test sets have 8000 patterns.}
\end{figure*}
\section{AODNN application examples.}
We confirm the feasibility and programmability of our AODNNs by building a fully-connected two-layer all optical neural network to recognize both handwritten digits and fashion images. The network structure and its optical layout are shown in Fig.\ref{fig:figure3}(a) and \ref{fig:figure3}(b). There are 144, 20, and 10 optical neurons in the input, hidden and output layer, respectively. A collimated control laser beam (marked with red color) with horizontal linear polarization, generated from a fiber output through lens L1 is shone on SLM1 that is placed on the back focal plane of lens L2. The SLM1 is divided into 144 sections and reflects the control light to produce 144 input nodes with controllable weights through the programmed phase gratings on each section (See \textbf{Supplementary Information S3 and Fig.S1-S3}). An aperture stop is placed on the front focal plane of L2 to filter away unwanted high-order diffraction modes. Then these 144 spatial nodes are imaged through the lens L3 on the SLM2. The combination of the SLM2 and the lens L4 performs the first linear matrix operation $\mathbf{W_1}$. The 20 outputs of this linear operation are then imaged on the cold atoms in MOT through a telescopic setup of the lenses L5 and L6. A quarter-wave plate (QWP1) converts the horizontal linear polarization into $\sigma^+$ circular polarization to meet the EIT requirement. The control beam transverse mode area inside the MOT is about 400 $\mu$m$^2$. A probe laser beam (marked with blue color, $\sigma^+$ circularly polarized after QWP2) from the fiber output is collimated by the lens L7 with a beam diameter 8 mm and is shone to the MOT with the opposite direction to the control beams. The measured 20 nonlinear activation functions are plotted in \textbf{Supplemental Information S4 and Fig.S4 }. The nonlinear functions of different nodes are not exactly the same since $OD$, $\gamma_{12}$ and probe beam intensity are spatially varied. The probe beam, after passing through the atomic medium spatially dressed by the control beams, contains the output spatial patterns of the 20 hidden neurons. Passing through QWP1, the probe beam nodes becomes vertically polarized and is reflected by the PBS. The 20 hidden neuron outputs are then imaged on the SLM3 through the lenses L6 and L8. The combination of SLM3 and L9 performs the second linear operation $\mathbf{W_2}$ and generates 10 output nodes recorded by a camera.

\begin{figure}[tb]
\includegraphics[width=\linewidth]{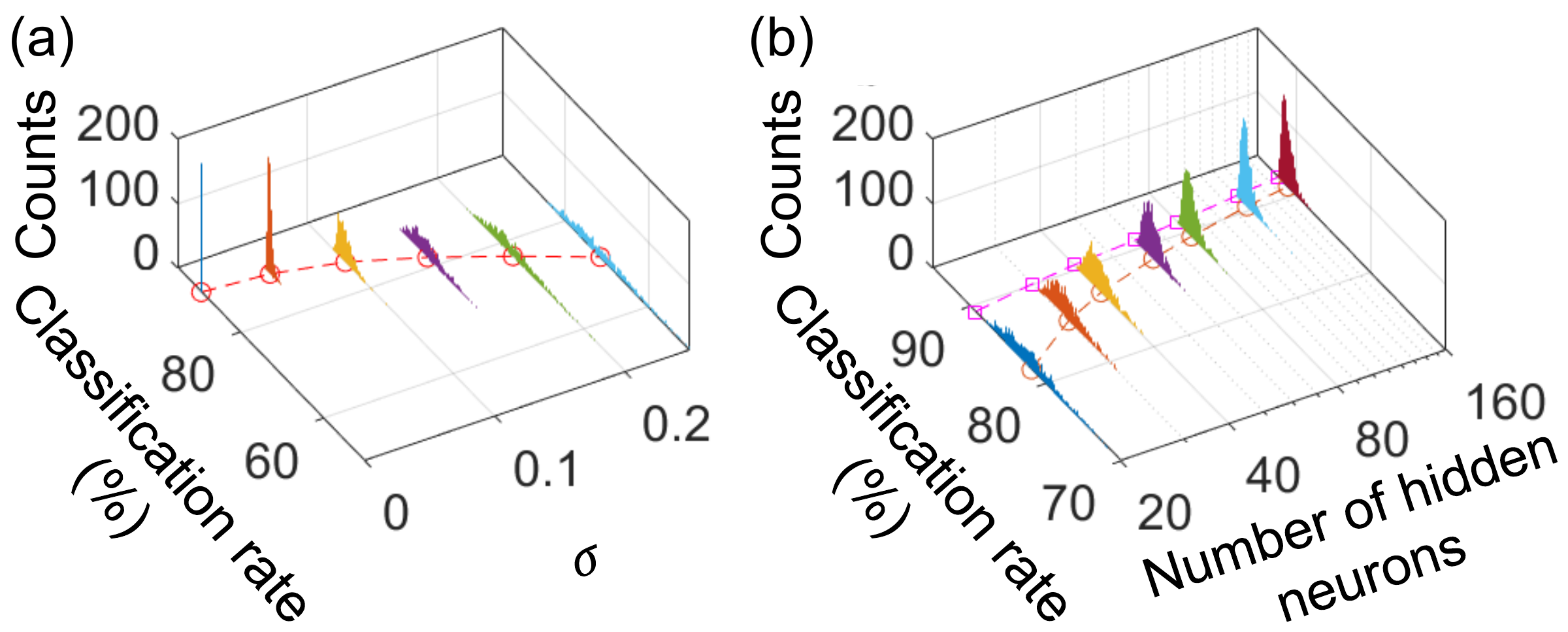}
\centering
\caption{\label{fig:figure5} \textbf{Classification rate of a two-layer AODNN simulation as a function of (a) standard deviation of random relative local errors and (b) the number of hidden neurons.} \textbf{a}, Numerical simulation of a two-layer AODNN by fixing 144 input neurons, 20 hidden neurons and 10 output neurons. Random relative errors sampled from Gaussian distribution $\mathcal{N}(0,\sigma^{2})$ are added to the weight matrix elements with $\sigma$ from 0 to 0.25. \textbf{b}, Numerical simulation of two-layer AODNNs by fixing 144 input neurons, 10 output neurons and varying the number of hidden neurons. Random relative errors sampled from Gaussian distribution $\mathcal{N}(0,\sigma^{2})$ is added to the weight matrix elements with $\sigma=0.15$. On both \textbf{a} and \textbf{b}, we perform different 1000 simulations with each combination of $\sigma$ and the number of hidden neurons. The histogram shows the classification rates distribution of 1000 simulations. The red circles show the average classification rate of 1000 simulations. The squares present classification rate of simulation without error. The two-layer AODNNs are trained with MNIST database. }
\end{figure}

We first test the ability of the AODNN for handwritten digit recognition. The initial training is done with a computer based neural network with the same network structure and nonlinear activation functions. We train the classifier using Modified National Institute of Standards and Technology (MNIST) handwritten digit database\cite{MNIST} (Fig.~\ref{fig:figure4}(a), compressed into images with 12$\times$12 resolution) and obtained the optimal parameters of the two linear matrix operations $\mathbf{W_1}$ and $\mathbf{W_2}$(See \textbf{Supplementary Materials MatrixElement.mat})
We achieve a classification rate of $81.8\%$, and the corresponding normalized confusion matrix is shown in Fig.~\ref{fig:figure4}(b), which is very close to the computer trained normalized confusion matrix in Fig.~\ref{fig:figure4}(c).
The AODNN hardware is programable. As a second example, we retrain the network with the fashion images\cite{xiao2017fashionmnist} (Fig.~\ref{fig:figure4}(d)) and configure the AODNN with completely different parameters (See \textbf{Supplementary Materials MatrixElement.mat}). We obtain a classification rate of $71.3\%$. The normalized confusion matrix of the AODNN (Fig.~\ref{fig:figure4}(e)) also agrees well with that of the computer trained neural network (Fig.~\ref{fig:figure4}(f)). 
To further demonstrate the ability of our AODNN, we use the computer to simulate the results with more hidden neurons due to the limited physical resources available in our lab. We first simulate the dependency of classification rate on the local errors that are added on each neuron randomly and independently. We performed 1000 different simulations with random relative local errors generated from Gaussian distribution $\mathcal{N}(0,\sigma^{2})$, where $\sigma$ is the standard deviation of relative local error. As shown in the Fig.~\ref{fig:figure5}(a), the classification rate indeed drops with local error as expected. However, the drop is relatively slow, 
and encouragingly the classification rate still can achieve 71.5\% for the large relative error 0.25. We choose the random relative error of standard deviation $\sigma=0.15$ to simulate the dependency of classification rate on the number of hidden neurons. As shown in Fig.~\ref{fig:figure5}(b), the classification rate increases with the number of hidden neurons, and the variance of the classification rate also becomes smaller for a large number of hidden neurons. It means that the effect of local random error become smaller and smaller in a larger-size AODNN. Due to the high consistency between our experiments and numerical simulations as proved on the recognition on both handwritten digit and fashion images above, we can expect similar behaviors in experiments when enough physical source is available. These results demonstrate that the drop of the classification rate induced by the local error can be easily compensated by increasing the number of hidden neurons.
In addition, we also numerically demonstrate that the usage of EIT nonlinear activation function can indeed improve the classification successful rate as shown in \textbf{supplementary Information S9 and Fig.S7}.

\section{Conclusion}
Through a systematical error measurement and analysis, we show that the AODNN based on SLM, Fourier optics and EIT nonlinear activation functions is scalable. We find that the errors remain local, random and independent, ($\propto\sqrt{N/S}$) and the error accumulation only follows square-root law of the network dimensions ($\propto\sqrt{MN/S}$),
and hence effect of these errors can be easily compensated by increasing the number of hidden neurons\cite{sung2015resiliency}.
We confirm the feasibility and programmability of AODNN with handwritten digits and fashion images recognitions and obtain comparable classification rates to computer based neural networks with the same structure. Although our experimental demonstrations take example of a two-layer AODNN architecture with only one hidden layer limited by our physical resource in lab, it can be extended into a much deeper AODNN. In next hidden layer, we can prepare atoms in state $\ket{2}$ in a second MOT, where the probe beams from the previous hidden layer serve as ``control" signals for the EIT nonlinearity and control beams become ``probe" outputs. By carefully designing the EIT parameters (such as atom number and ground state dephasing rates), a functional deeper AODNN with multiple hidden layers is achievable. For an AODNN, we can also insert optical repeaters (such as laser power amplifier) to compensate the passive losses in a very deep network. The transverse size can be simply extended with more SLMs and lenses. Our results suggest that a larger scale AODNN will be promising for light-speed optical computation and could be a powerful AI hardware.

\begin{acknowledgments}
Y. Zuo and Y. Zhao contributed equally to this work. Y. C. C. acknowledges the support from the Undergraduate Research Opportunities Program at the Hong Kong University of Science and Technology. J. L. acknowledges the support from the Hong Kong Research Grants Council (Projects No. 16306220).
\end{acknowledgments}

\appendix

\section{Gerchberg-Saxton (GSW) algorithm}
The weighted Gerchberg-Saxton (GSW) algorithm ~\cite{di2007computer} is a widely used algorithm. We developed an adaptive GSW iteration program to optimize the phase patterns of a SLM to obtain target weights in a linear transformation. To do so, we insert a flip mirror to the AODNN to reflect the light beams and image them into a camera. The difference between the measured intensity pattern and the target is feeded back to adjust the phase pattern of the SLM until desired weights with acceptable errors are achieved. More details are shown in \textbf{Supplementary Information S5 and Fig.S5}.

\section{Linear capacity of a single SLM-lens layer}
We use a SLM and a Fourier lens to perform a linear matrix operation. There are total $K$ pixels in the SLM and the area of each pixel is $d^2$. With total $M$ input optical neurons, each node occupies a rectangular area of $K d^2/M$. On the front focal plane of the Fourier lens with a focal length $f$, the corresponding Fourier transformed spot area is $4Mf^2\lambda^2/(Kd^2)$ (See \textbf{supplementary Information S6}). With the lens diameter $D$, the maximum number of nodes on the Fourier plane is $N_{max}=\frac{K\pi D^2d^2}{4Mf^2\lambda^2}$. Then the linear capacity is estimated as $C_L=M\times N_{max}=\frac{NA^2\pi Kd^2}{4\lambda^2}$, where $NA=D/(2f)$ is the numerical aperture of the lens.

\bibliography{AONN}

\end{document}